\documentclass[12pt]{article}
\usepackage{amsfonts,epsfig,latexsym}
\usepackage{pstricks,pst-node}
\catcode`\@=11 \@addtoreset{equation}{section}
\def\theequation{\arabic{equation}}
\def\theequation{\thesection\arabic{equation}}

\def\JHEP#1#2#3{{\it J. High Energy Phys.} {\bf#1} (19#2) #3}

\newcommand{\be}{\begin{equation}}
\newcommand{\ee}{\end{equation}}
\newcommand{\ben}{\begin{displaymath}}
\newcommand{\een}{\end{displaymath}}
\newcommand{\ba}{\begin{eqnarray}}
\newcommand{\ea}{\end{eqnarray}}

\newcommand{\bean}{\begin{eqnarray*}}
\newcommand{\eean}{\end{eqnarray*}}

\def\@normalsize{\@setsize\normalsize{15pt}\xiipt\@xiipt
\abovedisplayskip 14pt plus3pt minus3pt%
\belowdisplayskip \abovedisplayskip
\abovedisplayshortskip  \z@ plus3pt%
\belowdisplayshortskip  7pt plus3.5pt minus0pt}
\def\small{\@setsize\small{13.6pt}\xipt\@xipt
\abovedisplayskip 13pt plus3pt minus3pt%
\belowdisplayskip \abovedisplayskip
\abovedisplayshortskip  \z@ plus3pt%
\belowdisplayshortskip  7pt plus3.5pt minus0pt
\def\@listi{\parsep 4.5pt plus 2pt minus 1pt
\itemsep \parsep \topsep 9pt plus 3pt minus 3pt}}
\def\underline#1{\relax\ifmmode\@@underline#1\else
$\@@underline{\hbox{#1}}$\relax\fi} \@twosidetrue \relax
\catcode`@=12
\catcode`\@=11

\def\section{\@startsection{section}{1}{\z@}{3.5ex plus 1ex minus
.2ex}{2.3ex plus .2ex}{\large\bf}}
\def\thesection{\arabic{section}.}

\def\ps@headings{\def\@oddfoot{}\def\@evenfoot{}
\def\@oddhead{\hbox{}\hfill
\makebox[.5\textwidth]{\raggedright\ignorespaces --\thepage{}--
\hfill }}
\def\@evenhead{\@oddhead}
\def\subsectionmark##1{\markboth{##1}{}} }
\ps@headings \catcode`\@=12 \relax

\def\figcap{\section*{Figure Captions\markboth
{FIGURECAPTIONS}{FIGURECAPTIONS}}\list {Fig.
\arabic{enumi}:\hfill}{\settowidth\labelwidth{Fig. 999:}
\leftmargin\labelwidth\advance\leftmargin\labelsep\usecounter{enumi}}}
\relax
\def\tablecap{\section*{Table Captions\markboth
{TABLECAPTIONS}{TABLECAPTIONS}}\list {Table
\arabic{enumi}:\hfill}{\settowidth\labelwidth{Table 999:}
\leftmargin\labelwidth
\advance\leftmargin\labelsep\usecounter{enumi}}}
 \relax
\def\reflist{\section*{References\markboth
{REFLIST}{REFLIST}}\list
{[\arabic{enumi}]\hfill}{\settowidth\labelwidth{[999]}
\leftmargin\labelwidth
\advance\leftmargin\labelsep\usecounter{enumi}}}
 \relax
\catcode`\@=11
\def\marginnote#1{}
\newcount\hour
\newcount\minute
\newtoks\amorpm
\hour=\time\divide\hour by60 \minute=\time{\multiply\hour by60
\global\advance\minute by- \hour}
\edef\standardtime{{\ifnum\hour<12 \global\amorpm={am}%
\else\global\amorpm={pm}\advance\hour by-12 \fi \ifnum\hour=0
\hour=12 \fi
\number\hour:\ifnum\minute<100\fi\number\minute\the\amorpm}}
\edef\militarytime{\number\hour:\ifnum\minute<100\fi\number\minute}
\def\draftlabel#1{{\@bsphack\if@filesw {\let\thepage\relax
\xdef\@gtempa{\write\@auxout{\string
\newlabel{#1}{{\@currentlabel}{\thepage}}}}}\@gtempa
\if@nobreak \ifvmode\nobreak\fi\fi\fi\@esphack}
\gdef\@eqnlabel{#1}}
\def\@eqnlabel{}
\def\@vacuum{}
\def\draftmarginnote#1{\marginpar{\raggedright\scriptsize\tt#1}}
\def\draft{\oddsidemargin -.5truein
\def\@oddfoot{\sl preliminary draft \hfil
\rm\thepage\hfil\sl\today\quad\militarytime}
\let\@evenfoot\@oddfoot \overfullrule 3pt
\let\label=\draftlabel
\let\marginnote=\draftmarginnote
\def\@eqnnum{(\theequation)\rlap{\kern\marginparsep\tt\@eqnlabel}%
\global\let\@eqnlabel\@vacuum}  }
\def\preprint{\twocolumn\sloppy\flushbottom\parindent 1em
\leftmargini 2em\leftmarginv .5em\leftmarginvi .5em \oddsidemargin
-.5in    \evensidemargin -.5in \columnsep 15mm \footheight 0pt
\textwidth 250mmin      \topmargin  -.4in \headheight 12pt
\topskip .4in \textheight 175mm \footskip 0pt
\def\@oddhead{\thepage\hfil\addtocounter{page}{1}\thepage}
\let\@evenhead\@oddhead \def\@oddfoot{} \def\@evenfoot{}  }
\def\titlepage{\@restonecolfalse\if@twocolumn\@restonecoltrue\onecolumn
\else \newpage \fi \thispagestyle{empty}\c@page\z@
\def\thefootnote{\fnsymbol{footnote}} }
\def\endtitlepage{\if@restonecol\twocolumn \else  \fi
\def\thefootnote{\arabic{footnote}}
\setcounter{footnote}{0}}  
\catcode`@=12 \relax

\def\ps@headings{\def\@oddfoot{}\def\@evenfoot{}
\def\@oddhead{\hbox{}\hfill
\makebox[.5\textwidth]{\raggedright\ignorespaces --\thepage{}--
\hfill }}
\def\@evenhead{\@oddhead}
\def\subsectionmark##1{\markboth{##1}{}} }
\ps@headings \relax\def\firstpage#1#2#3#4#5#6{
\begin{document}
\begin{titlepage}
\nopagebreak
\title{\begin{flushright}
\vspace*{-1.8in}
{\normalsize CPTH RR 009.0105}\\[-6mm]
{\normalsize LPT-ORSAY 05/06}\\[-6mm]
{\normalsize hep-th/0502085}\\[-6mm]
\end{flushright}
\vfill {#3}}
\author{\large #4 \\[1.0cm] #5}
\maketitle \vskip -7mm \nopagebreak
\begin{abstract} {\noindent #6}
\end{abstract}
\vfill
\begin{flushleft}
\rule{16.1cm}{0.2mm}\\[-3mm]
$^{\dagger}${\small Unit{\'e} mixte du CNRS et de l'EP, UMR 7644.}\\[-3mm]
$^{\ddagger}${\small Unit{\'e} mixte du CNRS, UMR 8627.}\\
\end{flushleft}
\thispagestyle{empty}
\end{titlepage}}
\def\simlt{\stackrel{<}{{}_\sim}}
\def\simgt{\stackrel{>}{{}_\sim}}
\newcommand{\dal}{\raisebox{0.085cm} {\fbox{\rule{0cm}{0.07cm}\,}}}
\newcommand{\dt}{\partial_{\langle T\rangle}}
\newcommand{\dtbar}{\partial_{\langle\overline{T}\rangle}}
\newcommand{\al}{\alpha^{\prime}}
\newcommand{\mst}{M_{\scriptscriptstyle \!S}}
\newcommand{\mpl}{M_{\scriptscriptstyle \!P}}
\newcommand{\dv}{\int{\rm d}^4x\sqrt{g}}
\newcommand{\lv}{\left\langle}
\newcommand{\rv}{\right\rangle}
\newcommand{\ph}{\varphi}
\newcommand{\abar}{\overline{a}}
\newcommand{\sbar}{\,\overline{\! S}}
\newcommand{\xbar}{\,\overline{\! X}}
\newcommand{\fbar}{\,\overline{\! F}}
\newcommand{\zbar}{\overline{z}}
\newcommand{\dbar}{\,\overline{\!\partial}}
\newcommand{\tbar}{\overline{T}}
\newcommand{\taubar}{\overline{\tau}}
\newcommand{\ubar}{\overline{U}}
\newcommand{\ybar}{\overline{Y}}
\newcommand{\phb}{\overline{\varphi}}
\newcommand{\cm}{Commun.\ Math.\ Phys.~}
\newcommand{\prl}{Phys.\ Rev.\ Lett.~}
\newcommand{\pr}{Phys.\ Rev.\ D~}
\newcommand{\pl}{Phys.\ Lett.\ B~}
\newcommand{\ibar}{\overline{\imath}}
\newcommand{\jbar}{\overline{\jmath}}
\newcommand{\np}{Nucl.\ Phys.\ B~}
\newcommand{\F}{{\cal F}}
\renewcommand{\L}{{\cal L}}
\newcommand{\A}{{\cal A}}
\newcommand{\e}{{\rm e}}
\newcommand{\dslash}{{\not\!\partial}}
\newcommand{\gsi}{\,\raisebox{-0.13cm}{$\stackrel{\textstyle
      >}{\textstyle\sim}$}\,} 
\newcommand{\lsi}{\,\raisebox{-0.13cm}{$\stackrel{\textstyle <}{\textstyle\sim}$}\,}
\date{}
\firstpage{3118}{IC/95/34} {\large\bf Internal magnetic fields and
supersymmetry in orientifolds } { E. Dudas$^{\,a,b}$ and C.
Timirgaziu$^{\,b,a}$}
{\\[-3mm]
\normalsize\sl $^a$ Centre de Physique
Th{\'e}orique$^\dagger$, Ecole Polytechnique, F-91128 Palaiseau, France\\[-2mm]
\normalsize\sl $^b$  LPT$^\ddagger$, B{\^a}t. 210, Univ. de
Paris-Sud, F-91405 Orsay,
France\\[-3mm]}
{ Within the context of type I strings, we show the equivalence between BPS  
D9 branes with internal magnetic fluxes $H_i$ in the three torii and
non-BPS D3 branes with inverted internal magnetic fluxes $1/H_i$.  
 We then construct new supersymmetric examples of $Z_2 \times Z_2$
 orientifolds with discrete torsion which in the past had only
 non-supersymmetric solutions and emphasize the role of new twisted
 tadpole cancellation conditions, arising in the presence of magnetic
 fields, in order to get a consistent spectrum.
 In a second and independent part of the paper, we construct a new 
nine-dimensional type IIB orientifold with
Scherk-Schwarz deformation which has the peculiarity of introducing
a new type of non-BPS $O9$ planes and which contains as 
top branes a Scherk-Schwarz deformation of non-BPS D9 branes. 
The model contains charged D7 and D3 branes with a soft supersymmetry
breaking spectrum. 
} \break

\section{Introduction}

Internal magnetic fluxes in string theory first proved their
interesting features by providing the an explicit realization
of the non-linear Born-Infeld electrodynamics \cite{ft,acny}. 
One of its simplest consequences is that, since magnetic  fields
couple differently to particles of different spins, they naturally
break supersymmetry \cite{earlier}, a notoriously difficult problem   
in string theory. Complete string theory vacua with ${\cal N}=1$ 
supersymmetry or no supersymmetry at all were constructed some time ago 
in the T-dual picture
of branes intersecting at angles \cite{intersecting} or in the internal
magnetic fields description \cite{aads}. A collective effort over the
last years did lead to models of particle physics closer and closer
to the Standard Model or its minimal supersymmetric
extension, at the same time providing new ways of implementing inflation
in string theory \cite{inflation}. One of the new interesting features
of internal magnetic fluxes is that they induce RR charges and tensions
for branes of different dimensionalities \cite{li}. This fact, of
crucial importance for finding consistent string vacua 
\cite{aads,intersecting}, has also the welcome feature of allowing
radical rank reductions for the gauge group and of allowing new
supersymmetric orientifold vacua. 

Our present paper was trigerred by the curious but by now
well-established fact that some of the induced tensions on magnetized
D-branes can be negative. This interesting observation does raise the
hope of finding new supersymmetric solutions in situations where
this was regarded as impossible in the past, especially in models
featuring the phenomenon of ``brane supersymmetry breaking'' \cite{ads,aaads}. 
Indeed, this issue was recently raised in \cite{blt,ms} in a particular
class of constructions and \cite{ms} did display non-chiral supersymmetric
$Z_2 \times Z_2$ orientifold constructions in this framework, using
both BPS and non-BPS bulk D-branes with magnetic fluxes in their worldvolume.
The goal of the present paper is a more detailed understanding of the
conditions under which supersymmetric constructions do exist, 
to set the general consistency rules and to find the massless
spectrum for this new type of supersymmetric $Z_2 \times Z_2$ orientifold constructions 
with discrete torsion and branes at the orbifold fixed points.  

The organization of the paper is as follows. In Section 2 we show that
BPS and non-BPS D-branes in type I strings in the presence of internal
magnetic fluxes are fully equivalent at the string level by performing 
a simple mapping of the fluxes from the BPS to the non-BPS branes.  
In Section 3 we present general considerations and necessary conditions
for obtaining supersymmetric constructions with three internal magnetic
fields on an arbitrary number of stacks of D-branes, which forces upon
specific orbifold constructions.  
Section 4 presents our main new results containing new chiral supersymmetric
 $Z_2 \times Z_2$ orientifolds with discrete torsion with D-branes
at orbifold fixed points. In this case, we also show that the twisted RR 
tadpole conditions add new consistency constraints compared to the
corresponding orientifolds without magnetic fields. We present the
general pattern of the gauge group and of the massless spectrum and
give explicit examples. 
 Section 5, independent of the rest of the paper, contains a new
orientifold of Scherk-Schwarz type II string compactifications in
nine dimensions, which has the new feature of generating non-BPS
orientifold O9 planes of a new type. The top D9 branes in the model
are a soft Scherk-Schwarz deformation of non-BPS branes of type I
strings, whereas the model contains charged D7 and D3 branes with a soft
supersymmetry breaking spectrum.
  Finally, we present our conclusions and present in a short appendix
some useful formulae and conventions.

\section{Non-BPS versus BPS brane transmutation in type I strings}

The type I string has stable BPS D9 and D5 branes (also D1 which will play no
role in our discussion), whereas all its other Dp branes are
unstable. Of particular interest for our present discussion are, however,
the non-BPS D7 and D3 branes, which were constructed by Sen \cite{sen}
starting from Type II D-branes and consistently imposing the
orientifold projection. Taking into account that the type I orientifold
projection antisymmetrizes the RR 8-form and 4-form fields coupling to
D3 and D7 branes, the construction starts from an equal pair $M=N$ of
brane-antibrane pairs in Type IIB, which are interchanged by the
orientifold projection. The procedure creates a new stack of uncharged
branes, with gauge group $U(D)$, where $D=M=N$, obtained from the
original gauge group $U(M) \times U(N)$ by a suitable identification
of the Chan-Paton charges. The CFT description of a stack of non-BPS
D3 and D7 branes was presented in \cite{dms}.   
  
A simple and popular way of breaking partly or totally
supersymmetry is by adding magnetic fluxes on the BPS branes
\cite{earlier,intersecting,aads}. Considering for simplicity a factorizable internal space $T^2 \times
T^2 \times T^2$ of coordinates $w_i = x_i + i y_i$ and 
volumes $v_i$, we will denote the internal
magnetic fields on D9 branes pointing in the three different torii by $H_1,H_2,H_3$.
The three magnetic fields then satisfy a generalized version of the
Dirac quantization condition
\be
H_i = {m_i \over n_i v_i} \ , \label{t1}
\ee 
where $(m_i,n_i)$ are integers generalizing the Landau levels of a
particle in a magnetic field in quantum mechanics. By performing 3 T-dualities in the
coordinates $x_i$, one in
each compact torus $T_i^2$, we can trade the magnetic fluxes for D6-brane rotations  
\be
\tan \theta_i = H_i \ , \label{t2}
\ee
where $\theta_i$ is the angle in the torus $T_i^2$ made by the D6
brane with the coordinate $y_i$.  Arbitrary rotations in the three
tori break completely supersymmetry. For particular cases, however,
partial supersymmetry is restored. For example, $\theta_3=0$ and 
$\theta_1 \pm \theta_2=0$ leaves unbroken ${\cal N}=2$ supersymmetry
from a four-dimensional viewpoint, whereas 
\be
\theta_1 \pm \theta_2 \pm \theta_3 = 0 \label{t3} 
\ee
with all angles different from zero preserves  ${\cal N}=1$
supersymmetry in toroidal compactifications. By using (\ref{t1}), the conditions (\ref{t3})
translate into non-linear conditions for the internal magnetic fields,
which for the $(+,+)$ sign in (\ref{t3}), for example, is 
\be
H_1 + H_2 + H_3 = H_1 H_2 H_3 \ . \label{t4} 
\ee

The RR charges of various fields are given by the Wess-Zumino terms
\ba
&& {n_1 n_2 n_3 \over 2}  
\left( q_9 \int_{{\cal M}_{10}} C \wedge e^{s_1 H_1 + s_2H_2 + s_3H_3} + 
q_9 \int_{{\cal M}_{10}}  C \wedge e^{-s_1 H_1
- s_2 H_2 - s_3 H_3} \right) \ \nonumber \\
&& = n_1 n_2 n_3 q_9 \ \int_{{\cal M}_{10}} C_{10} + \nonumber \\  
&& n_1 n_2 n_3
q_9 \  \int_{{\cal M}_{10}} C_6 \wedge (s_1 s_2 H_1 \wedge H_2 + s_2 s_3 H_2 \wedge H_3 + 
s_1 s_2 H_1 \wedge
H_3 )  
\  \label{t5}
\ea
where $s_i/2 = \pm 1/2$, $i=1,2,3$ are the internal helicities of the (left or
right from the worldsheet point of view) fermions. 
The terms in (\ref{t5}) correspond in the effective action to the RR couplings 
\ba
&& n_1 n_2 n_3 q_9 \int_{{\cal M}_{10}}  \ C_{10} +  q_5 \left [
s_2 s_3 n_1 m_2 m_3  \int_{{\cal M}_{6}^{(1)}} \ C_6^{(1)}+
s_1 s_3 m_1 n_2 m_3  \int_{{\cal M}_{6}^{(2)}} \ C_6^{(2)}  \right. \nonumber \\
&& \left. + s_1 s_2 m_1 m_2 n_3   \int_{{\cal M}_{6}^{(3)}} \ C_6^{(3)} 
\right]
\  \label{t05}
\ea
where $q_9$ is the charge of one D9 brane, $q_5$ is the charge of one 
D5 brane and $C_6^{(i)}$ are six-form RR fields coupling to $D5_i$
branes with worldvolumes wrapping tori $T^2_i$.  
Therefore the brane acquires lower-dimensional charges, as if it were to
contain lower-dimensional branes \cite{li}. By using the quantization
conditions (\ref{t1}), these are precisely induced charges of the D5
branes type in the internal space which will modify tadpole
consistency conditions \cite{aads}.   

On the other hand, for arbitrary rotation angles the effective tension
of the BPS branes with fluxes cannot be understood in terms of the
original tension plus induced lower dimensional tensions. Indeed, the
tension of the magnetized D9 branes, corresponding in character
language to the brane coupling to the closed string field
$V_2O_2O_2O_2$ in the tree-level channel cylinder
amplitude, is
\be
T \ = \ |n_1 n_2 n_3| T_9 \ \sqrt{(1+H_1^2) (1+H_2^2)(1+H_3^2)} \ , \label{t6}
\ee 
where $T_9=q_9$ is the tension of a single D9 brane. 
Eq. (\ref{t6}), combined with the quantization condition (\ref{t1}), cannot be given
generally the same simple and elegant interpretation as the RR charge
(\ref{t5}). In the particular case where the supersymmetry condition is
satisfied, however, this becomes possible and matches precisely the RR
charge interpretation. For example, in the case  $\theta_3=0$ and 
$\theta_1 \pm \theta_2=0$, (\ref{t1}) gives the constraint
$(m_1/n_1v_1) = \pm (m_2/n_2 v_2)$ and the induced tension reads 
$T \ = \ T_9 \ (|n_1n_2| + { |m_1m_2| \over v_1v_2}) \ = \ |n_1n_2| \
T_9 \ + \  |m_1m_2| T_5$, corresponding in the effective theory
in the string frame to the terms
\be
 |n_1n_2| \ T_9 \int d^{10} x \ \sqrt{g} \ e ^{-\phi} \ +
 |m_1m_2| T_5 \int d^{6} x \ \sqrt{g} \ e ^{-\phi} \ , \label{t7} 
\ee 
where $T_5$ is the tension of a single D5 brane. In this case, the
induced tension is always positive and the magnetic fluxes can mimic
D5 brane or antibrane tensions and charges \cite{aads}.  
In the case of three rotations preserving ${\cal N}=1$ supersymmetry,
the situation is similar, with one important difference. 
For the $(+,+)$ sign in (\ref{t3}), for example and combining
(\ref{t4}) with (\ref{t6}), we find
\ba
&& T_{\rm eff} \ = \ |n_1 n_2 n_3 \ T_9 \ (1 - H_1 H_2 - H_2 H_3 - H_1 H_3 ) | \
= \ |n_1 n_2 n_3 \ T_9  \ \label{t8} \\ 
&&- T_5 \ ( n_1 m_2m_3 + \ m_1 n_2 m_3 +  
\ m_1m_2 n_3) \ sgn (n_1n_2n_3) | \ . \nonumber   
\ea   
Similarly, the
couplings of the three internal volume (Kahler) fields
$O_2 V_2 O_2 O_2$, $O_2 O_2 V_2 O_2$ and $O_2 O_2 O_2 V_2$,
obtained factorizing open string amplitudes in the tree-level closed
channel, using (\ref{t4}) can be shown to be
\ba
&& |n_1 n_2 n_3| T_9 \ (1-H_1^2) 
\sqrt{ (1+H_2^2)(1+H_3^2) \over (1+H_1^2)}
 =  |n_1 n_2 n_3| T_9  (1 + H_1 H_2 - H_2 H_3 + H_1 H_3 ) \ , 
\nonumber \\
&&  |n_1 n_2 n_3| T_9 \ (1-H_2^2) \sqrt{(1+H_1^2)(1+H_3^2) \over (1+H_2^2)}
 = |n_1 n_2 n_3| T_9  (1 + H_1 H_2 + H_2 H_3 - H_1 H_3 ) \ ,
\nonumber \\
&& |n_1 n_2 n_3| T_9 \ (1-H_3^2) 
\sqrt{(1+H_1^2) (1+H_2^2) \over (1+H_3^2)}
 =  |n_1 n_2 n_3| T_9  (1 - H_1 H_2 + H_2 H_3 + H_1 H_3 ) \ . 
\label{t08}
\ea

There is in this
case a correct match of the various tensions in (\ref{t8}) with
the various RR charges in (\ref{t5}), signaling spacetime
supersymmetry in the spectrum. 
The qualitative 
difference compared with the previous case (\ref{t7}) is that
now it is possible that one of the three induced tensions in
(\ref{t8}) be negative, as also emphasized recently in \cite{am}.
 
In orbifold constructions with internal fluxes, twisted RR tadpole
conditions can also induce new non-trivial constraints, as we will see in
explicitly for the case of $Z_2 \times Z_2$ orientifold with discrete torsion.  
 
We now turn to type I non-BPS branes with fluxes. Let us consider the
type I non-BPS D3 branes in a compact space setup and perform
six T-dualities. The result are the non-BPS D9 branes in
the type IIB orientifold $\Omega' = \Omega I_6 (-1)^{F_L}$, where the
orientifold projection identifies the D9-D${\bar 9}$ pairs of the type IIB
string. In this case, the conditions for having ${\cal
N}=1$ supersymmetry are\footnote{As for the BPS brane case, there are
several sign choices in toroidal compactifications. We choose for
definiteness the $(+,+,+)$ signs.}
\be
{\cal \phi}_1 +{\cal \phi}_2 + {\cal \phi}_3 = {\pi \over 2} \ , 
\label{t9}
\ee
where ${\cal \phi}_i$ are rotation angles for the non-BPS branes.
Analogously to (\ref{t4}), (\ref{t9}) translates into
a non-linear condition for the internal magnetic fields   
\be
{\cal H}_1 {\cal H}_2 + {\cal H}_2 {\cal H}_3 + {\cal H}_1 {\cal H}_3
= 1 \ , \label{t10}  
\ee
where ${\cal H}_i = m'_i/(n'_iv'_i)$ are magnetic fluxes added on the non-BPS D9 branes.      
Interestingly, even if the non-BPS D9 branes have no RR charges, magnetic
fluxes will induce lower dimensional ones, which can be found
starting from the D9-D${\bar 9}$ pairs of type IIB   
\ba
&& {n'_1 n'_2 n'_3 \over 2} \ \bigl( \ q_9  \int_{{\cal M}_{10}} C 
\wedge \ e^{s_1 {\cal H}_1 + s_2 {\cal H}_2 +
  s_3 {\cal H}_3} 
-  q_9  \int_{{\cal M}_{10}} C \wedge \ e^{- s_1 {\cal H}_1
- s_2 {\cal H}_2 - s_3 {\cal H}_3} \ \bigr) \ \nonumber \\
&& = n'_1 n'_2 n'_3 \ q_9 \ \int_{{\cal M}_{10}}  C_{8} \wedge 
\ (s_1 {\cal H}_1 + s_2 {\cal
  H}_2 + s_3 {\cal H}_3)   \nonumber \\
&& + \prod_{i=1}^3  (s_i n'_i)  \ q_9 \  \int_{{\cal M}_{10}}
C_4 \wedge {\cal H}_1 \wedge {\cal H}_2
\wedge {\cal H}_3   
\  \label{t11}
\ea
corresponding in the effective action to the RR couplings 
\ba
&& \prod_{i=1}^3 (s_i m'_i) q_3  \int_{{\cal M}_{4}} \ C_{4} +  \nonumber \\
&& q_7 \left [
s_1 m'_1 n'_2 n'_3 \int_{{\cal M}_{8}^{(1)}} \ C_8^{(1)}+
s_2 n'_1 m'_2 n'_3  \int_{{\cal M}_{8}^{(2)}} \ C_8^{(2)}+  
s_3 n'_1 n'_2 m'_3  \int_{{\cal M}_{8}^{(3)}} \ C_8^{(3)} 
\right]
\  \label{t12}
\ea
where $C_4$ is the four form coupling to D3 branes and  
$C_8^{(i)}$ are the eight-forms couplings to three different types
of D7 branes. The tension of non-BPS branes is similar 
to (\ref{t6}), after a rescaling, so that  
${\cal T}_9 = \sqrt{2} T_9$. Then in the supersymmetric case
(\ref{t9}), combining the analog of (\ref{t6}) with
(\ref{t9}), we find the non-BPS brane tension to be given by
\ba
&& {\cal T}  \ = \ | n'_1 n'_2 n'_3 {\cal T}_9 \ 
({\cal H}_1 + {\cal H}_2 + {\cal H}_3-
{\cal H}_1 {\cal H}_2 {\cal H}_3 ) | \  \nonumber \\
&& = \ | m'_1 m'_2 m'_3 \ {\cal T}_3 +  {\cal T}_7
( m'_1 n'_2 n'_3 \ + \
n'_1 m'_2 n'_3 \ + \ n'_1 n'_2 m'_3 ) |  
\ . \label{t13} 
\ea
This expression has the interesting property of having interpretation
in terms of purely lower dimensional tensions, while the original tension
has disappeared ! It also matches the RR charges (\ref{t11}),
justifying the presence of supersymmetry in the spectrum.
   
Since adding arbitrary magnetic fluxes erases the main
difference between BPS and non-BPS branes, one can wonder whether a precise mapping 
exists between BPS branes with fluxes $H_i$ and 
non-BPS branes with fluxes ${\cal H}_i$, a mapping
 $H_i \leftrightarrow {\cal H}_i $ under which 
(\ref{t4}) turns into (\ref{t10}) with the same spectrum
on both BPS and non-BPS branes. Interestingly enough, we find 
that such a transformation exists and is simply given by  
\be
H_i \quad \leftrightarrow \quad {1 \over {\cal H}_i} \ . \label{t14}
\ee 
The interpretation of (\ref{t14}) is that the BPS D9 branes with fluxes $(m_i,n_i)$
satisfying the condition (\ref{t3})-(\ref{t4}) are mapped after six
T-dualities $v_i \leftrightarrow 1/v'_i$
into non-BPS D3 branes with fluxes $(m'_i,n'_i)= (n_i,m_i)$,
satisfying  (\ref{t9})-(\ref{t10}). 
  The reason behind ({\ref{t14}) is actually easy to understand. Let
 us start with a stack of M non-BPS D3 branes in type I. The
 construction starting from type II branes and the spectrum,
pioneered in \cite{sen}, were described in detail in \cite{dms}.  
The open string amplitudes in the
presence of internal magnetic fields are modified to\footnote{For the
notations and conventions see. e.g. the last two review papers in
\cite{orientifolds} and the appendix of this paper.}    
\ba
&& {\cal A} =\int_0^{\infty} { dt \over t^{3}} \biggl\{ M {\bar M} \
{\tilde P}_1 {\tilde P}_2 {\tilde P}_3  (V_2 O_6 +O_2 V_6- S_2 S_6-C_2
C_6 )(0) {1 \over \eta^{6}(\tau)}+ \prod_{i=1}^3 (2n_im_i) \times \nonumber \\
&& \left[ {M^2 \over 2} \ 
(O_2 O_6 +V_2 V_6 - S_2 C_6 - C_2 S_6) (2 z'_i \tau) 
+ {{\bar M}^2 \over 2} \ 
(O_2 O_6 +V_2 V_6 - S_2 C_6 - C_2 S_6 ) (-2 z'_i \tau) \right] 
\nonumber \\
&& \times \prod_{i=1}^3 {i \eta \over
  \vartheta_1 (2 z'_i \tau)} \biggr\}  
{1 \over \eta^{2}(\tau)} \ , \nonumber \\
&& {\cal M} = \int_0^{\infty} {dt \over t^{3}} \prod_{i=1}^3 (2 m_i) \biggl\{
\left[ {M \over 2} ({\hat O}_2 {\hat O}_6+{\hat V}_2 {\hat
V}_6 )(2 z'_i\tau) +  {{\bar M} \over 2} 
({\hat O}_2 {\hat O}_6+{\hat V}_2 {\hat
V}_6) (-2 z'_i\tau) \right]   \  \nonumber \\
&& - \left[ {M \over 2} ({\hat S}_2 {\hat C}_6-{\hat C}_2 {\hat
S}_6 )(2 z'_i\tau) -  {{\bar M} \over 2} 
({\hat S}_2 {\hat C}_6-{\hat C}_2 {\hat
S}_6) (-2 z'_i\tau) \right] \bigg\} 
{1 \over \eta^{2}(\tau)} \prod_{i=1}^3 {i \eta \over
  \vartheta_1 (2 z'_i \tau)} \
 \ , \label{t15} 
\ea
where ${\cal \phi}_i \equiv \pi z'_i $ and where ${\tilde P}_i$ denote
"boosted" compactification lattices \cite{acny} obtained replacing the
Kaluza-Klein momenta ${\bf k}_i$ in the i$th$ torus by ${\bf k}_i
\rightarrow{\bf k}_i / \sqrt{n_i^2 + (m_i^2/v_i^2)} $  .
Then by using the identities (\ref{a1}) for the Jacobi functions, it is
straightforward to show that (\ref{t15}) transform precisely into the amplitudes of the
BPS D9 branes in type I with magnetic fluxes
\be
z_i = z'_i + {1 \over 2} \ , \label{t16} 
\ee
after 6 T-dualities. The mapping (\ref{t16}) implies (\ref{t14}) and
in particular explains, through the additional $\pi /2 $ rotations
in each torus, the change of the GSO projection in going from
the non-BPS branes in (\ref{t15}) to the BPS branes.   
Therefore, our results show that the examples proposed in \cite{blt,ms} with
  magnetic fluxes on non-BPS branes can be entirely described by
  appropriately modified fluxes on BPS branes. In 
  section 4 we describe new supersymmetric solutions of the $Z_2 \times
  Z_2$ orientifold with discrete torsion obtained using only BPS branes with
  internal magnetic fluxes.
    
\section{Supersymmetry and negative induced tensions~: general
  considerations }

At first sight, the internal fluxes on the BPS D9 branes induce D$5_i$ type tensions
(\ref{t8}), which are of two distinct types
\ba
&& (+,+,-) \ {\rm and \ permutations} \quad {\rm if} \quad H_1,H_2 > 0 \ , \ H_3
< 0 \ , \nonumber \\
&& (-,-,-) \quad {\rm if} \quad H_1, H_2,H_3 > 0 \ , \label{bsb1} 
\ea 
and the reversed solutions $ H_i \rightarrow - H_i$. 
The second case is incompatible, however, with the supersymmetry
condition (\ref{t3}). Notice that it is
not possible to generate only positive
induced  D$5_i$ type tensions. Let us consider in the following $p$
stacks of D9 branes $M_a$, $a = 1 \cdots p$. The ${\cal N}=1$ supersymmetry     
conditions (\ref{t4}) on each stack and the D9 and D5 tadpole
conditions in toroidal, $Z_2$ and $Z_2 \times Z_2$ orbifold
compactifications can be written compactly as
\ba
&&
m^{(a)}_1  n^{(a)}_2 n^{(a)}_3 v_2 v_3 + n^{(a)}_1  m^{(a)}_2
n^{(a)}_3 v_1 v_3 + n^{(a)}_1  n^{(a)}_2 m^{(a)}_3 v_1 v_2 =
\prod_{i=1}^3  m^{(a)}_i \quad , \quad {\rm for \ any \ a} \nonumber
\\
&& \sum_a M_a  n^{(a)}_1  n^{(a)}_2 n^{(a)}_3 = 16 \ , \nonumber \\
&&  \sum_a M_a  n^{(a)}_1  m^{(a)}_2 m^{(a)}_3 = - 16 \ \epsilon_1 \ ,
\nonumber \\
&&
 \sum_a M_a  m^{(a)}_1  n^{(a)}_2 m^{(a)}_3 = -16 \ \epsilon_2 \ ,
\nonumber \\
&&
\sum_a M_a  m^{(a)}_1  m^{(a)}_2 n^{(a)}_3 = - 16 \ \epsilon_3 \ ,
\label{bsb2}
\ea
where
\ba
&& (\epsilon_1, \epsilon_2, \epsilon_3) = (0, 0, 0) \quad {\rm in \ toroidal
  \ comp.} \ , \nonumber \\
&& (\epsilon_1, \epsilon_2, \epsilon_3) = (\pm 1, 0, 0) \quad {\rm in} \ Z_2 
 \  {\rm  comp.} \ , \nonumber \\
&& (\epsilon_1, \epsilon_2, \epsilon_3) = (\pm 1, \pm 1, \pm 1) \quad
{\rm in} \ Z_2 \times Z_2 \  
 {\rm  comp.}  \ . \label{bsb3}
\ea
Defining  $n^{(a)} \equiv  n^{(a)}_1 n^{(a)}_2 n^{(a)}_3$, the
tadpole conditions in (\ref{bsb2}) can also be written 
\ba
&& \sum_a M_a  n^{(a)} = 16 \ , \nonumber \\
&&  \sum_a M_a  n^{(a)} H^{(a)}_2 H^{(a)}_3 = - 16 \ {\epsilon_1 \over
  v_2 v_3} \ ,
\nonumber \\
&&
 \sum_a M_a  n^{(a)} H^{(a)}_1 H^{(a)}_3 = -16 \ {\epsilon_2 \over v_1
   v_3 } \ ,
\nonumber \\
&&
\sum_a M_a  n^{(a)} H^{(a)}_1 H^{(a)}_2 = - 16 \ {\epsilon_3 \over v_1
  v_2 } \ .
\label{bsb4}
\ea
Let us consider in the following the case with all $ n^{(a)}_i > 0$. 
Combining (\ref{bsb4}) with (\ref{t8}), we immediately find

i) if 
\be
1- \sum_{i < j} H_i^{(a)} H_j^{(a)} \ > \ 1 \ , \label{bsb5}
\ee
then 
\be
{\epsilon_1 \over   v_2 v_3} + {\epsilon_2 \over   v_1 v_3} +
{\epsilon_3 \over   v_1 v_2} \ > \ 0 \ . \label{bsb6}
\ee 

Therefore, in this case the toroidal compactifications\footnote{Supersymmetric toroidal examples were
recently provided in \cite{am}, in the case of non-zero off-diagonal 
fluxes between the internal tori. We do not consider generalized
constructions of this type here.} and the
$T^4/Z_2$ orientifold example with exotic O$5_{-}$ planes are excluded
(they are incompatible with supersymmetry), the $Z_2 \times Z_2 $ orientifold  
without discrete torsion $\epsilon_i = (1,1,1)$ is possible, whereas
for the other cases necessary (but not sufficient) conditions are
provided by
\ba
&& \epsilon_i = (1,1,-1) \ {\rm possible} \ {\rm if} \ v_1 + v_2 \ > \
v_3 \ , \nonumber \\
 && \epsilon_i = (1,-1,-1) \ {\rm possible} \ {\rm if} \ v_1 \ > \ v_2 +
v_3 \ . \label{bsb7} 
\ea 

ii) if 
\be
1- \sum_{i < j} H_i^{(a)} H_j^{(a)} \ < \ - 1 \ , \label{bsb8} 
\ee
then 
\be
{\epsilon_1 \over   v_2 v_3} + {\epsilon_2 \over   v_1 v_3} +
{\epsilon_3 \over   v_1 v_2} \ < \ - 2 \ . \label{bsb9}
\ee 
In this case the toroidal constructions are again impossible, as well
as the standard $T^4/Z_2$ orientifold with O$5_{+}$ planes and the 
$Z_2 \times Z_2 $ orientifold  
without discrete torsion $\epsilon_i = (1,1,1)$. For the other cases,
there are conditions analogous to (\ref{bsb7}) which we do not
explicitly display for brevity. 
 
\section{ Chiral supersymmetric $Z_2 \times Z_2$ models with discrete
torsion}

We now turn to four dimensional compactification on \mbox{$T^{2}\times
T^{2}\times T^{2}$} of the  Type-IIB theory, orbifolded by the
\mbox{$Z_{2}\times Z_{2}$} action generated by the identity (we will
call it ``$o$'') and the $\pi$ rotations $g:(+,-,-)$, $f:(-,+,-)$,
$h:(-,-,+)$, where the three entries within the parentheses refer to the
three internal tori, while ``$+$'' and ``$-$'' are the two group 
elements of $Z_{2}$.

There are several choices actually, depending on three signs $\epsilon_i = \pm 1$, where
$\epsilon_i=1$ typically signals the existence of $O5_{+}$ planes and
$\epsilon_i=-1$ that of  $O5_{-}$ planes. The different
possibilities are restricted by the condition
\be
\epsilon \quad = \quad \epsilon_1 \ \epsilon_2 \ \epsilon_3 \ . \label{f01}
\ee

The case  $\epsilon = 1$ in (\ref{f01}) defines models without discrete
torsion, whereas the case $\epsilon = -1$ in (\ref{f01}) defines models with discrete
torsion. The simplest model without discrete torsion  $\epsilon_i =
(1,1,1)$ was worked out in \cite{bl1,bl2} . It was proposed some time ago that
all the other models, having at least one $\epsilon_1 = -1$, have no 
supersymmetric solution. The reason advanced in
\cite{aaads} is that  $O5_{i,-}$ planes have positive tension and
positive charge and would ask for a supersymmetric solution for
compensating negative tension and charge,  impossible
to obtain by adding branes. Therefore RR tadpole conditions
ask for the introduction of $\overline{D5}_{i}$ antibranes and  
supersymmetry is necessarily broken. This conclusion, as emphasized
recently in \cite{ms},  should be revised
in models with three internal magnetic fields backgrounds, according
to the possibility of having one negative induced tension in
(\ref{t8}). In contrast to \cite{blt,ms}, however, we consider here
fractional D-branes, i.e. branes coupling to twisted sector fields.      
One possible advantage from a model
building point of view in these models compared to the (simpler) ones
without discrete torsion is that in all $Z_2 \times Z_2$ orientifold models without discrete
torsion constructed in the literature there are three chiral (super)fields in
the adjoint representation of the gauge group, an unwanted
feature for phenomenological applications. 
On the other hand, the bulk branes considered in \cite{ms} lead to non-chiral spectra, whereas
fractional branes allow in our constructions to get chiral spectra.  
The closed string amplitudes for all $Z_2 \times Z_2$ orientifolds are
described by the torus amplitude\footnote{For definitions of $Z_2 \times
Z_2$ characters, see e.g. \cite{aaads}.}
\ba 
\mathcal{T} &=& \int {d^2 \tau \over \tau_2^3} 
\frac{1}{4} \biggl\{ \Lambda_{1} \Lambda_{2} 
\Lambda_{3} |T_{oo}|^{2} + \Lambda_{1}
 |T_{og}|^{2} 
|\frac{4\eta^{2}}{\vartheta^{2}_{2}}|^{2}+ \Lambda_{2} 
|T_{of}|^{2}
 |\frac{4\eta^{2}}{\vartheta^{2}_{2}}|^{2} + \Lambda_{3}
|T_{oh}|^{2}  |\frac{4\eta^{2}}{\vartheta^{2}_{2}}|^{2}
  \nonumber\\
&&  + \Lambda_{1}
|T_{go}|^{2} \frac{4\eta^{2}}{\vartheta^{2}_{4}}|^{2} +
\Lambda_{1}
|T_{gg}|^{2} \frac{4\eta^{2}}{\vartheta^{2}_{3}}|^{2} 
+ |T_{fo}|^{2} \Lambda_{2}| \ \frac{4\eta^{2}}{\vartheta^{2}_{4}}|^{2}
+ |T_{ff}|^{2} \Lambda_{2}| \ \frac{4\eta^{2}}{\vartheta^{2}_{3}}|^{2}
 \nonumber\\
&& 
 + \Lambda_{3} |T_{ho}|^2 \frac{4\eta^{2}}{\vartheta^{2}_{4}}|^{2}
+ \Lambda_{3} |T_{hh}|^2 \frac{4\eta^{2}}{\vartheta^{2}_{3}}|^{2}
 \nonumber \\
&&  + \epsilon \ 
|\frac{8\eta^{3}}{\vartheta_{2}\vartheta_{3}\vartheta_{4}}|^{2}
( |T_{gh}|^{2}+|T_{gf}|^{2}+|T_{hg}|^{2}
+|T_{hf}|^{2}+|T_{fg}|^{2}+|T_{fh}|^{2} )  \biggr\} {1 \over |\eta|^4} 
 \ , \label{z1}
\ea
and the Klein bottle amplitude
\ba 
{\mathcal K} &=& \int_0^{\infty} {d \tau_2 \over \tau_2^3}
\frac{1}{8} \biggl\{ (P_{1}P_{2}P_{3} + P_{1} W_{2} W_{3}  + 
W_{1} P_{2} W_3+ W_{1} W_{2} P_3) T_{oo} + 
\nonumber\\ 
&&  32 \left( \frac{\eta}{\vartheta_{4}}\right)^{2}
 \left[ \epsilon_1 (P_{1}+\epsilon W_{1}) T_{go}
+ \epsilon_2 (P_{2}+\epsilon W_{2}) T_{fo} + 
\epsilon_3 (P_{3}+\epsilon W_{3}) T_{ho} \right]
 \biggr\} {1 \over \eta^2} \ . \label{z2}
\ea

The simplest orientifold model without discrete torsion $\epsilon_i
=(+,+,+) $ \cite{bl1,bl2} has $48$
chiral multiplets from the twisted sector, whereas the example
with $(+,-,-)$ has $16$ chiral multiplets and $32$ vector multiplets
from the twisted sector. For the models with discrete torsion, both
choices $(+,+,-)$ and $(-,-,-)$ yield a massless twisted closed string
spectrum composed of $48$ chiral multiplets.  
The crucial difference between the models without discrete torsion and
the models with discrete torsion is that in the first case the
twisted sector fields cannot couple to the D-branes and therefore, by
the open-closed duality, the orbifold has no action on the Chan-Paton
factors, whereas in the second case the branes can have couplings to
twisted fields which ask for new, twisted RR tadpole conditions.  
The case of interest for us, satisfying (\ref{f01}) is, up to permutations,
$(\epsilon_1,\epsilon_2,\epsilon_3) = (1,1,-1)$. In this case, we
have configurations of $O5_{1,+}$ planes,  $O5_{2,+}$ and
$O5_{3,-}$ planes. Due to the action of the orbifold operations on the
Chan-Paton (CP) factors, the appropriate CP parametrization for the
case under consideration for several stacks of fractional branes is
\ba
&&M_{a,o} = \ p_a + q_a + {\bar p}_a + {\bar q}_a \ , \  
M_{a,g} = i \ (p_a + q_a - {\bar p}_a - {\bar q}_a ) \ , \nonumber \\
&& M_{a,f} = i \ (p_a - q_a - {\bar p}_a + {\bar q}_a ) \ , \   
M_{a,h} = \ p_a - q_a + {\bar p}_a - {\bar q}_a  \ . \label{z3}
\ea 
In the corresponding models with discrete torsion but without the
internal magnetic fields the twisted tadpole conditions read 
\be
\sum_a M_{a,g} = \sum_a M_{a,f} = \sum_a M_{a,h} = 0 \ , \label{z04}  
\ee
which have the simple solution $p_a = q_a$ in order to cancel the
couplings to the twisted RR fields, giving gauge
groups of the form $\prod_a U(p_a)^2 $. In the case with internal
magnetic fields however, with the stack $a$ experiencing the fluxes
$(m^{(a)}_i, n^{(a)}_i)$, the conditions $p_a = q_a$, while still
necessary, are not sufficient anymore, since the new twisted RR tadpoles are
\ba
&& \sum_a (p_a + q_a) \ m_1^{(a)} \ = \ 0 \ , \nonumber \\ 
&&  \sum_a (p_a - q_a) \ m_2^{(a)} \ = \ 0 \ , \   \sum_a (p_a - q_a) \ m_3^{(a)} \ = \ 0 , 
 \label{z05}\
\ea
and therefore, in addition to $p_a = q_a$, we obtain   
the new condition 
\be
\sum_a (p_a + q_a) \ m_1^{(a)} \ = \ 0 . \label{z4}
\ee 
The reason behind this new condition is that the couplings of twisted
six-dimensional RR fields ${\cal C}_6$ (e.g. $S_2 C_2 O_2 O_2$) to the magnetized
D-branes are of the form
\ba
&& {n_1^{(a)} \over 2}  
\left(  (p_a+q_a) \int_{{\cal M}_{6}} {\cal C} \wedge \  e^{s_1
 H_1^{(a)}} -  ({\bar p}_a+{\bar q}_a)
\int_{{\cal M}_{6}}  {\cal C} \wedge \  e^{-s_1 H_1^{(a)}} \right) \ \nonumber \\
&& = {1 \over v_1} s_1 \ ( p_a + q_a + {\bar p}_a + {\bar q}_a ) \ m_1^{(a)} 
\int_{{\cal M}_{4}} {\cal C}_4    
\ .  \label{z5}
\ea
Similarly to the case of section 2, where unphysical RR couplings of the non-BPS branes
were turned into physical couplings by the internal magnetic fields,
unphysical couplings of twisted RR
fields to the D-branes are turned into physical couplings to twisted four-forms, 
leading to the new condition (\ref{z4}).

The cylinder amplitude, for models with discrete torsion containing
only magnetized D9 branes, is most
conveniently separated into several pieces. The strings propagating
from one stack $a$ of branes to the same or its image $a'$ are
described by
\ba
&& {\cal A}^{aa,aa'} =  \frac{1}{4} \int_0^{\infty} {dt \over t^3}
 \sum_a \biggl\{ |p_a +q_a|^2
 {\tilde P}_1 {\tilde
P}_2 {\tilde P}_3 (V_8-S_8) (0) {1 \over \eta^6}   \label{z6} \\
&& + \left[ |p_a +q_a|^2 {\tilde P}_1 T_{og} (0) + |p_a - q_a|^2 {\tilde P}_2
T_{of} (0) +  |p_a - q_a|^2 {\tilde P}_3 T_{oh} (0) \right] 
 ({2 \eta \over \vartheta_2})^2 \nonumber \\
&& + \prod_{i=1}^3 (2
m_i^{(a)} n_i^{(a)}) \left [ { (p_a +q_a)^2 \over 2} (V_8-S_8) (2 z_i^{(a)} \tau)
+ {({\bar p}_a +{\bar q}_a)^2
  \over 2} (V_8-S_8) (-2 z_i^{(a)} \tau) \right] \prod_{i=1}^3 {i \eta
  \over \vartheta_1 (2 z_i^{(a)} \tau)}  \nonumber \\
&& -
\left [ {  (p_a +q_a)^2 \over 2} T_{og} (2 z_i^{(a)} \tau) + { ({\bar p}_a +{\bar q}_a)^2  \over 2} T_{og}
(-2 z_i^{(a)} \tau) \right] \ \left( {2 i m_1^{(a)} n_1^{(a)} \eta
  \over \vartheta_1 (2 z_1^{(a)} \tau)} \right) \prod_{i=2,3} {2 \eta \over
  \vartheta_2 (2 z_i^{(a)} \tau)} \nonumber \\
&& -
\left[ {  (p_a - q_a)^2 \over 2 }  T_{of} (2 z_i^{(a)} \tau) +  { ({\bar p}_a -{\bar q}_a)^2  \over 2} T_{of}
(-2 z_i^{(a)} \tau) \right] \ \left( {2 i m_2^{(a)} n_2^{(a)} \eta
  \over \vartheta_1 (2 z_2^{(a)} \tau)} \right) \prod_{i=1,3} {2 \eta \over
  \vartheta_2 (2 z_i^{(a)} \tau)} \nonumber \\
&& +
\left [  {  (p_a - q_a)^2 \over 2 }  T_{oh} (2 z_i^{(a)} \tau) +  { ({\bar p}_a -{\bar q}_a)^2  \over 2} T_{oh}
(-2 z_i^{(a)} \tau) \right] \ \left( {2 i m_3^{(a)} n_3^{(a)} \eta
  \over \vartheta_1 (2 z_3^{(a)} \tau)} \right) \prod_{i=1,2} {2 \eta \over
  \vartheta_2 (2 z_i^{(a)} \tau)} \biggr\} {1 \over \eta^2} \ , \nonumber 
\ea
whereas the strings stretched between the stack $a$ and the stack $b$
(and their corresponding images) are described by
\ba
&& {\cal A}^{ab,ab'} =  \frac{1}{4}   \int_0^{\infty} {dt \over t^3} \sum_{a \not=b} 
\biggl\{ I^{ab} 
\left[ (p_a +q_a) ({\bar p}_b +{\bar q}_b) (V_8-S_8) ( z_i^{(ab)}
\tau)  + {\rm h.c.} \right] \prod_{i=1}^3 {i \eta
  \over \vartheta_1 (z_i^{(ab)} \tau)}   \nonumber \\
&&  +  I^{ab'} \left[ (p_a +q_a)(p_b +q_b)  (V_8-S_8) (z_i^{(ab')} \tau)
+ {\rm h.c.} \right] \prod_{i=1}^3 {i \eta
  \over \vartheta_1 (z_i^{(ab')} \tau)} \label{z7}  \\
&& +  \left[ I^{ab}_1  (p_a +q_a)  ({\bar p}_b +{\bar q}_b) \  
T_{og} (z_i^{(ab)} \tau) + {\rm h.c.} \right]  \left( {i  \eta
  \over \vartheta_1 (z_1^{(ab)} \tau)} \right) \prod_{i=2,3} {2 \eta \over
  \vartheta_2 (z_i^{(ab)} \tau)}  \nonumber \\
&& -
\left [ I^{ab'}_1 (p_a +q_a)  (p_b +q_b) \ T_{og} (z_i^{(ab')} \tau) + {\rm h.c.}
\right] \ \left( {i \eta
  \over \vartheta_1 (z_1^{(ab')} \tau)} \right) \prod_{i=2,3} {2 \eta \over
  \vartheta_2 (z_i^{(ab')} \tau)} \nonumber \\
&& +  \left[ I^{ab}_2  (p_a -q_a)  ({\bar p}_b -{\bar q}_b) \  
T_{of} (z_i^{(ab)} \tau) + {\rm h.c.} \right]  \left( {i  \eta
  \over \vartheta_1 (z_2^{(ab)} \tau)} \right) \prod_{i=1,3} {2 \eta \over
  \vartheta_2 (z_i^{(ab)} \tau)}  \nonumber \\
&& -
\left [ I^{ab'}_2 (p_a -q_a)  (p_b -q_b) \ T_{of} (z_i^{(ab')} \tau) + {\rm h.c.}
\right] \ \left( {i \eta
  \over \vartheta_1 (z_2^{(ab')} \tau)} \right) \prod_{i=1,3} {2 \eta \over
  \vartheta_2 (z_i^{(ab')} \tau)} \nonumber \\
&& +  \left[ I^{ab}_3  (p_a -q_a)  ({\bar p}_b -{\bar q}_b) \  
T_{oh} (z_i^{(ab)} \tau) + {\rm h.c.} \right]  \left( {i  \eta
  \over \vartheta_1 (z_3^{(ab)} \tau)} \right) \prod_{i=1,2} {2 \eta \over
  \vartheta_2 (z_i^{(ab)} \tau)}  \nonumber \\
&& +
\left [ I^{ab'}_3 (p_a -q_a)  (p_b -q_b) \ T_{oh} (z_i^{(ab')} \tau) + {\rm h.c.}
\right] \ \left( {i \eta
  \over \vartheta_1 (z_3^{(ab')} \tau)} \right) \prod_{i=1,2} {2 \eta \over
  \vartheta_2 (z_i^{(ab')} \tau)} \biggr\} {1 \over \eta^2} \ , \nonumber 
\ea
where we defined the intersection numbers of the magnetized D9 branes of the stacks
$a$ and $b$ ($b'$) in the i$th$ torus
\ba
&&I^{ab}_i = m^{(a)}_i n^{(b)}_i - n^{(a)}_i m^{(b)}_i \ , \
I^{ab'}_i = m^{(a)}_i n^{(b)}_i + n^{(a)}_i m^{(b)}_i \ , \nonumber \\
&& I^{ab} = \prod_{i=1}^3 I^{ab}_i \ , \ I^{ab'} = \prod_{i=1}^3
I^{ab'}_i \ , \label{z8}
\ea
and the effective fluxes on strings with one end on the stack $a$ and
the other end on the stack $b$ ($b'$) 
\be
z_i^{(ab)} = z_i^{(a)} - z_i^{(b)} \ , \
z_i^{(ab')} = z_i^{(a)} + z_i^{(b)} \ . \label{z9} 
\ee
The Mobius amplitude is
\ba
&& {\cal M} = - \int_0^{\infty} {dt \over t^3}
\biggl\{ \prod_{i=1}^3 ( m_i^{(a)} )  
\left[ (p_a+q_a) ({\hat V}_8-{\hat S}_8) (2 z_i \tau) + ({\bar p}_a + {\bar
  q}_a)  ({\hat V}_8-
{\hat S}_8) (-2 z_i^{(a)} \tau) \right]
 \prod_{i=1}^3 {i {\hat \eta}  \over {\hat \vartheta_1} (2 z_i \tau)} \nonumber \\
&& - \epsilon_1 \left[ (p_a+q_a) {\hat T}_{og} (2 z_i^{(a)} \tau) +
({\bar p}_a +{\bar q}_a)  {\hat T}_{og} (-2 z_i^{(a)} \tau) \right]  
\left( {i m_1^{(a)} {\hat \eta}  
\over {\hat \vartheta_1} (2 z_1^{(a)} \tau)} \right)
 \prod_{i=2,3} {n_i^{(a)} {\hat \eta} \over  {\hat \vartheta_2} (2
   z_i^{(a)} \tau)} \nonumber \\
&& - \epsilon_2 \left[ (p_a+q_a) {\hat T}_{of} (2 z_i^{(a)} \tau) +
({\bar p}_a + {\bar q}_a) {\hat T}_{of} (-2 z_i^{(a)} \tau) \right]  
\left( {i m_2^{(a)} {\hat \eta}  \over 
{\hat \vartheta_1} (2 z_2^{(a)} \tau)} \right)
 \prod_{i=1,3} {n_i^{(a)} {\hat \eta} \over  {\hat \vartheta_2} (2
   z_i ^{(a)} \tau)} \nonumber \\
&& - \epsilon_3 \left[ (p_a+q_a) {\hat T}_{oh} (2 z_i^{(a)} \tau) +
({\bar p}_a + {\bar q}_a) {\hat T}_{oh} (-2 z_i^{(a)} \tau) \right]  
\left( {i m_3^{(a)} {\hat \eta}  
\over {\hat \vartheta_1} (2 z_3^{(a)} \tau)}\right)
 \prod_{i=1,2} {n_i^{(a)}{\hat \eta} \over  {\hat \vartheta_2} 
(2 z_i^{(a)} \tau)} \ \biggr\} {1 \over \eta^2} \ . \label{z10}
\ea
Transforming the various amplitudes
(\ref{z2}),(\ref{z6}),(\ref{z7}),(\ref{z10}) 
in the tree-level closed channel and factorizing the resulting
amplitudes, we obtain
the untwisted RR tadpole cancellation conditions
\ba
&& \sum_a (p_a + q_a) \ n_1^{(a)}  n_2^{(a)} n_3^{(a)}  =  16 \quad ,  \nonumber \\
&& \sum_a (p_a + q_a) \  n_1^{(a)}  m_2^{(a)} m_3^{(a)}  \ = \ - 16 \ \epsilon_1 \ , \nonumber \\
&&  \sum_a (p_a + q_a) \  m_1^{(a)} n_2^{(a)} m_3^{(a)} \ =
\ - 16 \ \epsilon_2 \ , \nonumber \\
&& \sum_a (p_a + q_a) \  m_1^{(a)}  m_2^{(a)} n_3^{(a)} \ = \ - 16 \ \epsilon_3 \ , \label{z11}   
\ea 
to be supplemented precisely by the twisted RR condition (\ref{z4}).
We are interested in the following in supersymmetric
models. Consequently the fluxes on the magnetized D9 branes will have
to satisfy the supersymmetric condition (\ref{t4}) for each stack. 

In order to display generically the massless spectrum in this class of
models, let us define
\be
I^{aO} = 8 \ \left( m_1^{(a)}  m_2^{(a)} m_3^{(a)} -
\epsilon_1 m_1^{(a)} n_2^{(a)} n_3^{(a)}
- \epsilon_2 n_1^{(a)}  m_2^{(a)} n_3^{(a)} -
\epsilon_3 n_1^{(a)}  n_2^{(a)} m_3^{(a)} \right) \ . \label{z12}
\ee
Then the massless spectrum is based on the gauge group
$\prod_a {\bf U(p_a)} \otimes {\bf U(q_a)}$, with $p_a = q_a$ and
chiral (super)fields in the representations
\ba
&&
{\rm {\bf Multiplicity}} \qquad \qquad \qquad \qquad \qquad \qquad
\qquad \qquad \qquad {\rm {\bf Representation}} \  \nonumber \\
&&
1 \qquad \qquad \qquad \qquad \qquad \qquad \qquad \qquad \qquad \qquad
\qquad ({\bf p_a,{\bar q_a}}) + ({\bf {\bar p_a}, q_a}) 
\ , \nonumber \\
&&
 {1 \over 8} \left( I^{aa'} + I^{aO} -4 I^{aa'}_1 - 4 I^{aa'}_2
+ 4 I^{aa'}_3  \right) \quad \quad ({\bf {p_a(p_a-1) \over 2},1})
+  ({\bf 1, {q_a(q_a-1) \over 2}}) \ , \nonumber \\
&&
 {1 \over 8} \left( I^{aa'} - I^{aO} -4 I^{aa'}_1 - 4 I^{aa'}_2
+ 4 I^{aa'}_3  \right) \quad \quad ({\bf {p_a(p_a+1) \over 2},1})
+  ({\bf 1, {q_a(q_a+1) \over 2}}) \ , \nonumber \\
&&
  {1 \over 4} \left( I^{aa'} -4 I^{aa'}_1 + 4 I^{aa'}_2
- 4 I^{aa'}_3  \right)  \qquad \quad  \quad ({\bf p_a, q_a}) \ , \nonumber \\
&&
  {1 \over 4} \left( I^{ab'} -4 I^{ab'}_1 - 4 I^{ab'}_2
+ 4 I^{ab'}_3  \right)  \qquad \quad  \quad ({\bf p_a, p_b}) +  
({\bf q_a, q_b}) \ , \nonumber \\
&&
  {1 \over 4} \left( I^{ab'} -4 I^{ab'}_1 + 4 I^{ab'}_2
- 4 I^{ab'}_3  \right)  \qquad \quad  \quad ({\bf p_a, q_b}) \ , \nonumber \\
&&
  {1 \over 4} \left( I^{ab} + 4 I^{ab}_1 + 4 I^{ab}_2
+ 4 I^{ab}_3  \right)  \qquad \qquad  \qquad ({\bf p_a, {\bar p_b}}) +  
({\bf q_a, {\bar q_b}}) \ , \nonumber \\
&&
  {1 \over 4} \left( I^{ab} + 4 I^{ab}_1 - 4 I^{ab}_2
- 4 I^{ab}_3  \right)  \qquad \qquad  \qquad ({\bf p_a, {\bar q_b}})  \ , \label{z13}
\ea  
where above $a \not= b$ in order to avoid the overcounting of states.
It is a straightforward exercise to show that the irreducible anomalies 
$SU(p_a)^3$ and $SU(q_a)^3$ with the
spectrum (\ref{z13}) cancel precisely when the untwisted  (\ref{z11}) and the twisted
 (\ref{z4}) RR cancellation conditions are satisfied. The other gauge anomalies are taken
care by the generalized version of the four-dimensional Green-Schwarz
mechanism \cite{ggs}. 
\subsection{Explicit models} 

We can now provide examples of $Z_2 \times Z_2$ orientifolds with discrete
torsion containing only magnetized D9 branes and no D5 branes. 
The first class of examples are those in which the twisted tadpole 
condition (\ref{z4}) is simply satisfied
starting with a pair of magnetized  stacks containing equal numbers of
D9 branes with opposite magnetic fluxes.
An explicit example in this case is $M_1=M_2=8$ and internal magnetic fluxes
\ba
&& (m^{(1)}_i, n^{(1)}_i) \ = \ (1,1) , (1,1) , (-1,1) \ , \nonumber \\
&& (m^{(2)}_i, n^{(2)}_i) \ = \ (-1,1) , (-1,1) , (1,1)  \ . \label{e1}
\ea
Supersymmetry on each stack has actually several possible solutions in
terms of the compact volumes. One possible solution corresponds to
$(v_1,v_2,v_3) = (3,2,1)$ with the corresponding magnetic fluxes on the
two stacks equal to $H^{(1)} = (1/3,1/2,-1)$ and  $H^{(2)} =
(-1/3,-1/2,1)$, respectively.
The gauge group of the model is ${\bf U(4)}^2 \otimes {\bf U(4)}^2 $, where we 
wrote separately the gauge factors ${\bf U(4)}^2$ coming from
the first and from the second stack. The spectrum contains chiral multiplets in 
$ ({\bf 4,{\bar 4},1,1}) + ({\bf {\bar 4}, 4,1,1})+  ({\bf 1,1, 4,{\bar
4}}) + ({\bf 1,1, {\bar 4}, 4}) + 
8 \times [ ({\bf 6,1,1,1}) + ({\bf 1,6,1,1}) + 
({\bf 1,1,{\bar 6},1})
+ ({\bf 1,1,1,{\bar 6}}) ]$. 
In this example the twisted tadpole condition (\ref{z4}) is simply satisfied
starting with a pair of stacks with opposite magnetic fluxes.
It is easy to construct other similar examples of this type. We
did indeed construct examples with gauge group ${\bf U(2)}^2 \otimes 
{\bf U(2)}^2$, ${\bf U(4)}^2 \otimes {\bf U(1)}^2$, and a third one based on
the gauge group ${\bf U(2)}^2 \otimes {\bf U(1)}^2$. 

Our second class of examples correspond to models in which the twisted tadpole 
condition (\ref{z4}) is satisfied starting with two stacks of different 
numbers of magnetized D9 branes and  different compensating fluxes. 
An explicit example in this class is 
$M_1=8$, $ M_2=4$, with internal magnetic fluxes
\ba
&& (m^{(1)}_i, n^{(1)}_i) \ = \ (1,1) , (1,1) , (-1,1) \ , \nonumber \\
&& (m^{(2)}_i, n^{(2)}_i) \ = \ (-2,2) , (-1,1) , (1,1)  \ . \label{e2}
\ea
A supersymmetric solution for the internal volumes and internal magnetic
fields is the same as in the previous example. 
The gauge group of the model consists of two copies of the
gauge group ${\bf U(4)}^2 \otimes {\bf U(2)}^2$. 
The chiral spectrum of the
model is $ 8 \times [ ({\bf 6,1,1,1}) + ({\bf 1,6,1,1})] +
2 \times [ ({\bf 1,1,3,1})+({\bf 1,1,1,3})] + 12 \times 
({\bf 1,1,2,2})$, together with $36$
${\bf SU(4)}^2$ and ${\bf SU(2)}^2$ singlets charged under various 
${\bf U(1)}$ factors. \\

{\bf A chiral model with a Standard Model type gauge group} \\

Phenomenologically more interesting models can be constructed out of
three stacks of magnetized D9 branes. An explicit example we found is
based on three stacks containing six, four and two branes,
respectively, with $M_1=6$, $M_2=4$ and $M_3=2$.   
The fluxes on the three stacks are given by 
\ba
&& (m^{(1)}_i, n^{(1)}_i) \ = \ (1,1) , (1,1) , (-1,1) \ , \nonumber \\
&& (m^{(2)}_i, n^{(2)}_i) \ = \ (-1,1) , (-2,2) , (1,1) \ , \nonumber \\ 
&&  (m^{(3)}_i, n^{(3)}_i) \ = \ (-1,1) , (-1,1) , (1,1) \ , \label{e3}
\ea 
which correspond to supersymmetric fluxes with similar solutions for
the internal volumes as in the previous examples. The resulting gauge group 
is $[{\bf U(3)} \otimes {\bf U(2)} \otimes {\bf U(1)}]_1 \otimes
[{\bf U(3)} \otimes {\bf U(2)} \otimes
{\bf U(1)}]_2 $  and the massless spectrum contains
\ba
&&
{\rm {\bf Multiplicity}} \qquad \qquad \qquad \qquad \qquad \qquad
\qquad \qquad \qquad {\rm {\bf Representation}} \  \nonumber \\
&&
1 \qquad \qquad \qquad \qquad \qquad \qquad \qquad \qquad \qquad
\qquad ({\bf 3,1,1;{\bar 3},1~,1}) +({\bf {\bar 3},1,1~;3,1~,1}) + 
\nonumber \\
&& \qquad \qquad \qquad \qquad \qquad \qquad
\qquad 2 \times ({\bf 1,2~, 1~; 1,2~,1}) + ({\bf 1,1, 1~;1~, 1,1}) +  
({\bf 1,1, 1~;1~, 1,1})    
\ , \nonumber \\
&&
8  \quad \quad  \qquad \qquad \qquad \qquad \qquad \qquad
\qquad \qquad
({\bf 3,1~, 1~; 1 , 1,1}) + ({\bf 1,1,1~; 3,1~,1})
 \ , \nonumber \\
&&
4  \qquad \qquad \qquad \qquad \qquad \qquad \qquad \qquad \quad \quad ({\bf {\bar 3},1, 1~;1~,2,1}) +({\bf 1,2~,1; {\bar 3},1,1})
 \ , \nonumber \\
&&
12  \qquad \qquad \qquad \qquad \qquad \qquad
\qquad \qquad \quad ({\bf 1,2~,1~;1~,1,1 }) + 
({\bf 1,1,~1~; 1,2~,1 }) \ , \nonumber \\
&&
2  \quad  \qquad \qquad \qquad \qquad \qquad
\qquad \qquad \qquad  ({\bf 1,3~,1~;1~,1,1 }) +({\bf 1,1~,1~;1~,3,1 })  \ , \nonumber \\
&&
36  \quad  \qquad \qquad \qquad \qquad \qquad
\qquad \qquad \qquad ({\bf 1,1,1~;1,1,1}) \ ,  \label{e4}
\ea  
where the $38$ singlets have various $U(1)$ charges not explicitly
displayed in (\ref{e4}). The model has therefore two copies of the 
Standard Model gauge group with four generations of quarks and leptons,
eight Higgs multiplets plus two exotic states in the symmetric 
representations of ${\bf U(2)}_i$.     
The (non-chiral) states in the byfundamentals
of ${\bf U(3)_1} \otimes {\bf U(3)}_2$ , ${\bf U(2)_1} \otimes {\bf
U(2)}_2$, ${\bf U(1)_1} \otimes {\bf U(1)}_2$ in the first two lines of 
 (\ref{e4}), if given a vev, would break the gauge group to the diagonal
Standard Model gauge group 
${\bf U(3)} \otimes {\bf U(2)} \otimes {\bf U(1)}$. An alternative
channel breaks the gauge symmetry down to ${\bf U(3)}_1 
\otimes {\bf U(2)}_2 \otimes {\bf U(1)}_1$.  
The gauge symmetry breaking generates masses for various states. 
It is beyond the
scope of the present paper to study in detail the resulting models, but
it is encouraging to find in a rather simple way the Standard Model
gauge group and the representations which correspond to the quarks,
leptons and Higgs multiplets. \\

{\bf A chiral model in the $Z_2 \times Z_2$ orientifold
without discrete torsion} \\

Let us consider again the four dimensional compactification on $(T^2)^3$ of
type IIB theory orbifolded by the group $Z_2\times Z_2$ and without
discrete torsion ($\epsilon_i=1, \ i=1,2,3$). Let us add three stacks of 
magnetized D9 branes with $M_1=8, \ M_2=4, \ M_3=2$ and 32 $D5_1$ 
branes without magnetic fluxes.  The supersymmetric magnetic fluxes are given by:

  \ba (m_i^{(1)},n_i^{(1)})&=&(-1,1),(1,1),(1,1) \ , \nonumber\\
(m_i^{(2)},n_i^{(2)})&=&(-1,1),(1,1),(1,1) \ , \nonumber\\
(m_i^{(3)},n_i^{(3)})&=&(-2,2),(1,1,),(1,1) \ . \ea

 The gauge group of this model is $ [{\bf U(4)}]_{9_1}\times 
[{\bf U(2)}]_{9_2} \times
[{\bf U(1)}]_{9_3}\times [{\bf USp(16)}]_5 $.
The chiral spectrum contains, in addition to three chirals in the
adjoint of every gauge factor, Weyl fermions in the $D9_i\ - \ D9_j$
intersections in the following representations:
\ba 
9_1\ - \ 9_1\ &:& \ 8 \times ({\bf 6,1,1,1}) \quad , \quad 
9_{2,3}\ - \ 9_{2,3}\ \quad : \quad
8 \times ({\bf 1,1,1,1}) \ , \nonumber \\ 
9_1\ - \ 9_2\ &:& \ 8 \times
({\bf 4,2,1,1}) \quad , \quad
9_1\ - \ 9_3\ \quad : \quad 16 \times({\bf 4,1,1,1}) \ , \nonumber\\
9_2\ - \ 9_3\ &:& \ 16 \times ({\bf 1,2,1,1}) \  
\ea
and fermions in the $D9_i\ - \ D5$ intersections:
\ba 9_1\ - \ 5\ &:&\  ({\bf {\bar 4},1,1,{16}})+
({\bf {\bar 4},1,1,{\bar 16}}) \ , \nonumber\\
9_2\ - \ 5\ &:& \
({\bf 1,2,1,{16}})+
({\bf 1,2,1,{\bar 16}}) \ , \nonumber\\
9_3\ - \ 5\ &:& \ 2\times[({\bf 1,1,1, {16}})+({\bf 1,1,1,{\bar 16}})]
\ .
\ea
The $9_1-9_3$ states, if given an appropriate vev, would break 
${\bf U(4)} \rightarrow {\bf U(3)}$ leading to a standard model gauge group with
eight generations. In order for the D5 gauge group to play the role 
of a hidden sector, the $D9_i-D5$ states have to be given a mass by 
adding, for example, appropriate Wilson lines.

\section{A new orientifold of Scherk-Schwarz compactifications
and non-BPS orientifold planes}

The non-BPS D3 brane of Type I maps, after six T-dualities, into a D9
brane in the Type IIB orientifold defined by the orientifold
projection $\Omega' = \Omega I_6 (-1)^{F_L}$, where $I_6$ denotes six
parities in the internal space and $(-1)^{F_L}$ is
the left spacetime fermion number. In this section we will construct a
new orientifold model in nine dimensions in which 
such branes appear as D9 branes wrapped on the circle, with a
Scherk-Schwarz type deformation of their spectrum. The model
contains also a new type of non-BPS orientifold planes which have
different couplings than the known non-BPS D branes in type II 
orientifolds.     
We believe that this model, which is the fourth type of nine-dimensional
orientifold of Scherk-Schwarz type II strings, following the three
previous ones \cite{ads1}, \cite{dm}, is also the last possible
construction in nine dimensions.
The construction starts from the torus amplitude of Scherk-Schwarz
compactifications \cite{scherkschwarz,ss}
\ba
&& {\cal T} = \int {d^2 \tau \over \tau_2^6} \ 
\left[ (|V_8|^2 +|S_8|^2) \Lambda_{m,2n} - (V_8 {\bar S}_8 + S_8
{\bar V}_8) \Lambda_{m+1/2,2n} \right. \nonumber \\
&&+ \left. (|O_8|^2 +|C_8|^2) \Lambda_{m,2n+1}- (O_8 {\bar C}_8 + C_8
{\bar O}_8) \Lambda_{m+1/2,2n+1} \right] \ {1 \over |\eta|^{16}} 
\ , \label{o1} 
\ea
where the notations are defined, for example, in the last two
references in \cite{orientifolds}. The first orientifold
projection of Scherk-Schwarz type II strings \cite{bd,ads1} is the
standard left-right exchange $\Omega_1 = \Omega$, the second one
\cite{ads1} is based on  $\Omega_2 = \Omega I_1$, where $I_1$ is the parity 
in the Scherk-Schwarz coordinate and generates a supersymmetry
breaking perpendicular
to the brane worldvolume, whereas the third one \cite{dm}, which has
the virtue of eliminating the closed  string tachyon (and also the open string
tachyon, in the sense of requiring a net number of 16 D8 branes by the RR
tadpole conditions),    
is based on $\Omega_3 = \Omega I_1 (-1)^{f_L}$, where  $(-1)^{f_L}$ is
the left worldsheet fermion number.

The present construction starts from the orientifold projection
$\Omega_4 = \Omega \ \delta \ (-1)^{F_L}$, where $\delta$ is the shift
$X_9 \rightarrow X_9 + \pi R$. This is indeed a symmetry of the
theory, provided that the shift and $(-1)^{F_L}$ are combined together.
The Klein bottle amplitude   
\be
{\cal K} = { 1 \over 2} \ \int_0^{\infty} {d \tau_2} \ (V_8+S_8) \
(-1)^m P_m \ {1 \over \eta^8 (2 i \tau_2)}
\ \label{o2}
\ee
has the peculiarity of symmetrizing the RR sector by projecting out the
two form, whereas keeping the zero form and the four form. The BPS
branes of this orientifold are therefore D3 and D7 branes, whereas all
the other branes, including the top D9 branes we will describe in a
moment, are non-BPS. The consistency of the construction can also be
checked by performing the S-transformation into the tree-level closed
channel     
\be
{\tilde {\cal K}} = {2^5 \over 2} \ \int_0^{\infty} dl \ (O_8-
C_8) \ W_{2n+1} \ {1 \over \eta^8 (il)} \ , \label{o3}
\ee
which reveal the presence of a new type of non-BPS O-planes, with no
tension and RR charge, coupling only to the closed string tachyon and
to a massive RR 10 form.  
The model does not demand by RR tadpole conditions the addition of D9
branes. However, they can be added consistently with the string
constraints~: particle interpretation and open-closed string duality. 
 The open string amplitudes are 
\ba
&& {\cal A} =\int_0^{\infty} { dt \over t^{11/2}} \left[ N {\bar N} \ (V_8 P_m - S_8 P_{m+1/2}) 
+ {N^2+{\bar N}^2 \over 2} \ (O_8 P_m - C_8 P_{m+1/2}) \right] {1 \over \eta^{8}(it/2)}  \ , \nonumber \\
&& {\cal M} = \int_0^{\infty} {dt \over t^{11/2}}  (-1)^m \left[ \pm {(N+{\bar N}) \over 2} \
  P_m \ {\hat O}_8  
    \pm {(N-{\bar N}) \over 2} \ P_{m+1/2} \ {\hat C}_8 \right] {1
    \over {\hat \eta}^{8}(it/2+1/2)} \ \label{o4} 
\ea
and define consistent non-BPS D9 branes with gauge group 
${\bf U(N)}$. 
The four different possible signs in the Mobius amplitude define
the four different signs of the O9 couplings $(\pm, \pm)$ to the
would-be closed tachyon and RR 10 form. Let us concentrate for
simplicity on the  $(-, +)$ sign. Then the spectrum of the model
also contains open tachyons in the antisymmetric representations
${\bf N(N-1)/2} + {\bf {\bar N}({\bar N}-1)/2}$ and 
(Majorana in nine dimensions)
fermions in ${\bf N(N-1)/2} + {\bf {\bar N}({\bar N}+1)/2}$. Since the
fermions
have KK masses shifted by $1/2$, the spectrum
can clearly be interpreted as a Scherk-Schwarz deformation of
certain non-BPS branes with the same spectrum as the non-BPS D7 
branes in the type I  
strings. The notable difference in the present case, however,
is that the single brane case $N=1$ is stable, since the open string
tachyons disappear. In the case of the single D7 brane in Type I
string, the same phenomenon happens but the D7 brane is still unstable
due to the tachyons coming from the D7-D9 interactions.   
 
Similarly, the choice  $(+, -)$ gives the Scherk-Schwarz
deformed spectrum of the non-BPS D3 brane of Type I strings. 
In this case, the D9 brane is unstable due to the open string tachyon. 

Since the RR sector contains a zero-form, a four-form and an eight-form,
the model contains charged D3 and D7 branes. Let us, for concreteness,
discuss the case of D7 branes. The D7-D7 and D7-O9 
string amplitudes in the open channel 
are given by
\ba
&& {\cal A}_{77} = {M^2 \over 2} \int_0^{\infty} \ {dt \over t^{9/2}} 
\{ (V_6 O_2 + O_2 V_6) P_m - (S_6 S_2 + C_6 C_2) P_{m+1/2}  \} \ {1
  \over \eta^8} \ , \label{o5} \\
&& {\cal M}_{79} = - {M \over 2} \int_0^{\infty} \ {dt \over t^{9/2}}
\{ ({\hat O_6} {\hat V_2} -{\hat V_6}
{\hat O_2}) (-1)^m P_m + ({\hat S_6} {\hat S_2} -{\hat C_6}
{\hat C_2}) (-1)^m P_{m+1/2} \} \ {1
  \over \eta^8} \ . \nonumber     
\ea      
The gauge group for  $M$ coincident D7 branes is therefore ${\bf
  USp(M)}$ and 
the open spectrum includes two massless scalars in the antisymmetric
representation ${\bf M(M+1)/2}$, while the fermions are massive due to the
Scherk-Schwarz deformation. The whole open string spectrum (\ref{o5})
is manifestly a soft Scherk-Schwarz deformation of a supersymmetric
spectrum, whereas the closed string spectrum, due to the orientifold
projection leading to the Klein bottle (\ref{o2}), is manifestly
non-supersymmetric.     

The amplitudes in the closed, tree-level channel 
\ba
&& {\tilde {\cal A}}_{77} = {M^2 \over 2^5} \int_0^{\infty} {dl \over l} 
\{ (V_6 O_2 + O_2 V_6-S_6 S_2 - C_6 C_2 ) W_{2n} + 
(O_6 O_2 + V_6 V_2- S_6 C_2 - C_6 S_2) W_{2n+1}  \} {1
  \over \eta^8} \ , \nonumber \\
&& {\tilde {\cal M}}_{79} = {M }  \int_0^{\infty} {dl \over l}
\{ ({\hat O_6} {\hat O_2} +{\hat V_6}
{\hat V_2}) - (-1)^n ({\hat S_6} {\hat C_2} -{\hat C_6}
{\hat S_2}) \} \ W_{2n+1} {1 \over \eta^8 (il +1/2)} \ , \label{o6}     
\ea 
reveal couplings of the charged D7 branes with the dilaton and the
RR eight-form and also couplings to the would-be tachyon and to the
massive RR eight-form coming from the "twisted" sector.  

A similar analysis for $D$ coincident charged D3 branes gives a gauge
group ${\bf SO(D)}$.  

The peculiarity of the present model is the nature of 
supersymmetry breaking in the closed and in the open sector, in the
case with O9 and charged D7 branes. Whereas the D7-D7 brane interactions
are clearly a soft deformation of supersymmetric ones, for the closed
spectrum and the D7-O9 interactions, due to the orientifold projection 
$\Omega_4$, reflected 
in the Klein bottle amplitude, the softness manifests itself differently. 
Level by level supersymmetry seems
to be badly broken, but the breaking becomes soft when including the whole
spectrum. Indeed, the Klein and Mobius amplitudes actually vanish
in the $R \rightarrow \infty$ limit, as appropriate for a soft
supersymmetry breaking spectrum, displaying a kind of zig-zag supersymmetry 
providing cancellations between different mass levels.  
  
\section{Conclusions}

We showed the complete equivalence between BPS and non-BPS D-branes with appropriately
mapped internal magnetic fluxes in type I orientifold models. We
discussed some necessary conditions
obtained by requiring that one supersymmetry be preserved and the
tadpole conditions to be fulfilled, excluding
in particular toroidal and some orbifold constructions of this type.  

Our main goal in this paper was to reconsider some class of models featuring 
the phenomenon of ``brane supersymmetry breaking '' \cite{aaads}, in which exotic
O-planes of positive tension and charge did force the introduction of antibranes,
which did break supersymmetry.   
As anticipated in \cite{ms} and explicitly showed in Section 4, this conclusion can be avoided for
particular models with magnetized D9 branes and
negatively induced tension in one internal torus. Whereas the non-chiral constructions in 
\cite{ms} used bulk branes, we did construct explicit chiral
supersymmetric examples with branes at orbifold 
fixed points (fractional branes).
A novelty here is the appearance of new constraints from twisted RR
tadpole conditions due to internal magnetic fields which generate
physical four-dimensional couplings of D-branes to twisted RR fields. 
If our explicit examples are not fully realistic, one could reasonably expect that 
phenomenologically
quasi-realistic models along these lines can be constructed,
eventually combining the present constructions with recent approaches
to moduli stabilization \cite{fluxes}. Since in most of our constructions all
tadpole conditions were satisfied with only magnetized D9 branes, it is also plausible
that some of these constructions have $Z_2 \times Z_2$ heterotic duals \cite{alon}. 

Finally, we did present a new Scherk-Schwarz orientifold of type II
strings (in addition to the three already existing ones in the literature)
which involves a new type of non-BPS
O9 orientifold planes, coupling to the closed string tachyon. One of the possible future lines
of investigation in such models is related to the role of the  couplings to
the O-planes of the would-be closed tachyon for the tachyon
dynamics. Indeed, in the regime where the corresponding
closed string scalar becomes more and more tachyonic, its couplings to the O-planes
become more and more important and cannot be neglected. The soft nature of supersymmetry
breaking in this new orientifold construction is realized in an
interesting way, with the O9 plane
disappearing in the limit of supersymmetry restoration.   
\section{Appendix}
A useful Riemann identity for Jacobi functions is 
\ba
&& \sum_{\alpha,\beta=0,1/2}  \eta_{\alpha,\beta} \
\vartheta [{\alpha \atop \beta }] (z) \ \prod_{i=1}^3  \  
\vartheta[{\alpha \atop {\beta + v_i}}] = 
\nonumber \\
&& - 2 \ \vartheta_1 (-{z \over 2}) \ \vartheta_1 ({z -v_1+v_2+v_3 
\over 2}) \ \vartheta_1 ({z +v_1-v_2+v_3 \over 2}) \ \vartheta_1 
({z +v_1+v_2 -v_3 \over 2})
\ , \label{a1}
\ea
valid for $v_1+v_2+v_3 = 0$.

A notation used very often in the text is 
\be
(V_8-S_8) (z_i) \ \equiv \ \sum_{\alpha \beta} \eta_{\alpha \beta}
\ \vartheta [{\alpha \atop \beta }] (0) \ \prod_{i=1}^3  \  
\vartheta[{\alpha \atop {\beta - z_i}}] \ , \label{a2}
\ee
with similar notations for other parts of the amplitudes with open string propagations.

In all string amplitudes written in the text, we did not explicitly
write the contributions, proportional to $1/(4 \pi^2
\alpha')^{d/2}$ coming from the integral   
over the non-compact momenta, where $d$ is the number of non-compact
dimensions.

\setcounter{footnote}{0}
\section*{Acknowledgments}
This work is supported in part by the CNRS PICS no. 2530, INTAS grant 
03-51-6346, the RTN grants MRTN-CT-2004-503369, MRTN-CT-2004-005104 and  
by a European Union Excellence Grant, MEXT-CT-2003-509661.
We wish to thank C. Angelantonj, I. Antoniadis, R. Blumenhagen,
J. Mourad, G. Pradisi and A. Sagnotti for useful discussions.
ED would like to acknowledge the hospitality of the theory group of
CERN during the completion of this work.  


\end{document}